\documentclass[magazine]{IEEEtran}
\usepackage{color}
\usepackage{verbatim}
\usepackage{amsfonts}
\usepackage{amssymb}
\usepackage{stfloats}
\usepackage{cite}
\usepackage{graphicx}
\usepackage{psfrag}
\usepackage{subfigure}
\usepackage{amsmath}
\usepackage{array}
\usepackage{epstopdf}
\usepackage{authblk}
\usepackage{graphicx} 
\usepackage{amsthm} 
\usepackage{lipsum}
\usepackage{verbatim} 
\usepackage{authblk}
\usepackage{mathtools}
\usepackage{cuted}
\usepackage[lined,boxed,ruled]{algorithm2e}
\usepackage{algpseudocode}
\usepackage{framed} 
\usepackage{subfigure}
\usepackage{soul}
\usepackage{bm}
\usepackage{setspace}

\def\qed{$\Box$}

\def\phi{\varphi}

\def\({\left(}
\def\){\right)}

\def\b0{{\mathbf{0}}}

\title{Semantic Data Sourcing for 6G Edge Intelligence}

\author{Kaibin~Huang, Qiao~Lan, Zhiyan~Liu, and Lin Yang
\thanks{K. Huang, Q. Lan and Z. Liu are with Dept. of EEE, The University of Hong Kong, Hong Kong SAR. L. Yang is with Huawei Noah's Ark Lab, China. Contact: K. Huang (huangkb@eee.hku.hk).}}

\makeatletter
\newcommand{\removelatexerror}{\let\@latex@error\@gobble}
\makeatother

\IEEEoverridecommandlockouts

\begin{document}

\maketitle
\begin{abstract}
    As a new function of 6G networks, edge intelligence refers to the ubiquitous deployment of machine learning and \emph{artificial intelligence} (AI) algorithms at the network edge to empower many emerging applications ranging from sensing to auto-pilot. To support relevant use cases, including sensing, edge learning, and edge inference, all require transmission of high-dimensional data or AI models over the air. To overcome the bottleneck, we propose a novel framework of \emph{SEMantic DAta Sourcing} (SEMDAS) for locating semantically matched data sources to efficiently enable edge-intelligence operations. The comprehensive framework comprises new architecture, protocol, semantic matching techniques, and design principles for task-oriented wireless techniques. As the key component of SEMDAS, we discuss a set of machine learning based semantic matching techniques targeting different edge-intelligence use cases. Moreover, for designing task-oriented wireless techniques, we discuss different tradeoffs in SEMDAS systems, propose the new concept of joint semantics-and-channel matching, and point to a number of research opportunities. The SEMDAS framework not only overcomes the said communication bottleneck but also addresses other networking issues including long-distance transmission, sparse connectivity, high-speed mobility, link disruptions, and security. In addition, experimental results using a real dataset are presented to demonstrate the performance gain of SEMDAS.
\end{abstract}

\section{Introduction}
\label{sec: introduction}
The \emph{sixth-generation} (6G) mobile networks are expected to be \emph{artificial intelligence} (AI) native, featuring the ubiquitous deployment of machine learning and AI algorithms at the network edge~\cite{HuaweiWhitepaper}. On the other hand, data have replaced fuel to become the most valuable resource in the world~\cite{EconomistPaper}. Mobile data are being generated at an exponentially growing rate that is expected to increase by three-fold to reach 324 EB/month by 2028~\cite{EricssonWhitepaper}. The 6G edge intelligence will provide a platform for continuous distillation of AI to support many \emph{Internet-of-Everything} (IoE) applications ranging from sensing to auto-driving to industrial automation. Compared with “cloud AI”, edge intelligence has the advantages of efficient processing of mobile data, preserving user privacy, reducing network traffic, and providing ultra-low-latency access~\cite{Zhu2020CM}. The main challenge in implementing edge intelligence is that the wireless transmission of high-dimensional data or AI model parameters creates a communication bottleneck. To overcome the bottleneck, we propose in this article a novel framework of \emph{SEMantic DAta Sourcing} (SEMDAS) for locating semantically-matched data sources to efficiently enable edge-intelligence operations.

\subsection{5G Connectivity-Centric Networking}
5G’s key innovation lies in providing an infrastructure that supports heterogeneous types of services and applications. Specifically, to meet different requirements in rate, reliability, and latency, three types of connectivity have been defined including ultra-reliable-low-latency communications, massive machine-type communication, and enhanced mobile broadband. 5G communication techniques have been largely designed using the rate-centric approach that is rooted in Shannon’s information theory. In this approach, data are essentially treated as a sequence of bits that should be transported from a source to a destination reliably and quickly. As suggested by Warren Weaver in his 1953 paper~\cite{Weaver1953ETC}, communications should transcend merely solving this technical problem to address the semantic issue, namely the \emph{accuracy in communicating meanings of messages and the effectiveness of their use at the destination to execute a specific task.} One would argue that many mobile applications do have semantic awareness such as recommendations and advertising. However, the applications are add-ons to the network and interface only with its Application Layer. In other words, they are decoupled from the radio access layers where wireless techniques are deployed. The traditional \emph{computation-communication separation approach} for designing wireless techniques is believed to be sub-optimal in terms of \emph{end-to-end} (E2E) performance. As a result, the existing semantics-agnostic techniques lack the maximum efficiencies needed to cope with the exponentially growing mobile data and population of edge devices and achieve faster-than-human (i.e., 0.1 milliseconds) latency. Therefore, 6G researchers advocate the new \emph{computation-communication integration approach} that tightly couples communication, computing, sensing, and control in designing next-generation wireless techniques~\cite{HuaweiWhitepaper}.

\subsection{6G Semantic Communications}
As a 6G paradigm, \emph{Semantic Communications} (SemCom) refers to the computation-communication integrated designs that optimize the E2E performance metric of semantic accuracy or the overlapping metric of task effectiveness. In contrast with the traditional opaque-data transmission, the semantics-awareness can be exploited to reduce communication overhead by avoiding transmitting information lacking relevance and to improve the radio-resource utilization efficiency (e.g., prioritizing packets in resource allocation based on their data content). Two basic operations in a SemCom system are \emph{semantic encoding}, which compresses messages while retaining their semantics or task utility, and \emph{channel encoding} for ensuring reliability in the presence of channel distortion. In a popular approach known as joint source-channel coding, these two operations are enabled using separate neural network models, which are jointly trained for an E2E task such as edge inference or sensing~\cite{Jankowski2021JSAC}. For implementation on a layering network architecture, SemCom involves deep coupling of the top Application layer and bottom radio access layers~\cite{Qin2022arxiv,Lan2021JCIN}. To illustrate its advantage, consider the transmission of the message ``Einstein’s son, **** ****** ********, was a professor of ********* ************ at the University of ********************." with erroneous letters marked using ``*". It could have been discarded as an unreliable message in the Physical layer but can be reliably recovered in the Application layer using a knowledge graph to be ``Einstein’s son, Hans Albert Einstein, was a professor of hydraulic engineering at the University of California, Berkeley." The rise of edge intelligence provides a platform for SemCom where powerful tools including AI algorithms, knowledge graphs, and data analytics can automate and streamline the system operations. Furthermore, given an AI-empowered receiver robust against perturbation, the graceful degradation of inference performance with a decreasing number of transmitted features of source-data samples can also allow received messages to be useful even in the event of packet loss. 
\begin{figure*}[t]
    \centering
    \includegraphics[width=0.6\textwidth]{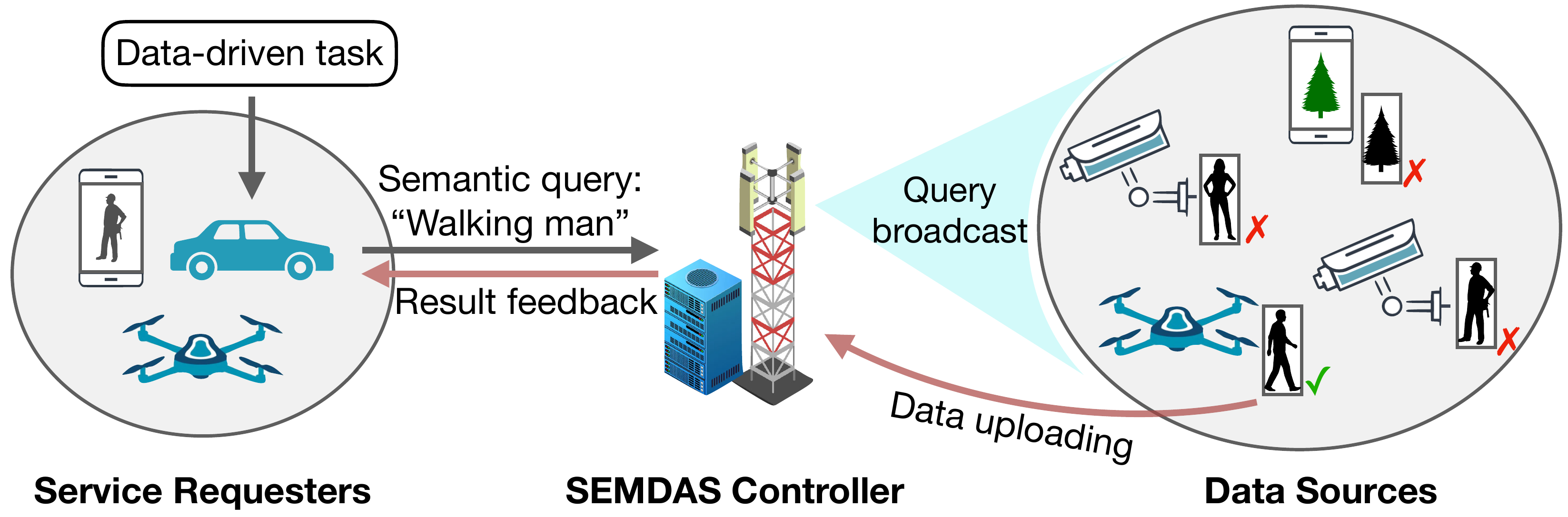}
    \caption{The SEMDAS network architecture. }
    \label{fig:overview}
\end{figure*}

\subsection{Proposed Semantic Data Sourcing for 6G Edge Intelligence}
Aiming at efficient implementations of edge intelligence on a SemCom platform, we propose a novel framework called \emph{SEMantic DAta Sourcing} (SEMDAS). The goal of SEMDAS is to locate among many data sources a subset that can provide semantically relevant data to enable communication-efficient edge learning or inference. To this end, the SEMDAS protocol essentially involves a service requester transmitting a query, which characterizes a specific edge-intelligence task, to a controller that searches for matched data sources. Then the matched sources are connected to the requester to provide a computing service or their data.  As the key component of SEMDAS, a set of semantic matching techniques are proposed for three representative use cases of edge intelligence — {Internet-of-Thing} (IoT) sensing, edge learning, and inference. They build on suitably chosen existing learning algorithms including semantic embedding, uncertainty evaluation, distribution matching, and autoencoder.

The proposed SEMDAS framework has three main advantages. First, SEMDAS overcomes the communication bottleneck of edge intelligence by avoiding unnecessary transmission of semantically irrelevant data. Second, without requiring E2E connections, SEMDAS exploits the existence of multiple semantically similar sources to address networking issues including long-distance transmission, sparse connectivity, high-speed mobility, and link disruptions. Last, data source verification via semantics checking by SEMDAS controller provides a mechanism to ensure security and data integrity. 

We further propose new principles for designing task-oriented wireless techniques for SEMDAS. One principle lies in system optimization based on the tradeoffs between query size, data overhead, privacy, and semantic matching accuracy. The other principle is joint semantics-channel matching. We discuss how the new tradeoffs lead to new designs of task-oriented wireless techniques including multi-access, over-the-air computing, radio resource management, and beamforming.


\section{Overview of Semantic Data Sourcing}
\label{sec: overview}
Although SEMDAS applies to a broader range of applications, we focus on those of edge intelligence. The network architecture, protocol, and advantages of SEMDAS are described separately in the following sub-sections. 

\subsection{SEMDAS Network Architecture}
The architecture, as illustrated in Fig.~\ref{fig:overview}, comprises the following key components.
\begin{itemize}
    \item \textbf{Service Requester:} The node (either a device or a server) performs a data-driven task. To this end, it sends a query that comprises descriptors to the SEMDAS controller to ask for semantically-matched data that helps effective task execution.
    \item \textbf{SEMDAS Controller:} The node coordinates the SEMDAS process, serves as the interface between data sources and requester, and implements the security function, and manages mobility.
    \item \textbf{Semantic Data Sources:} Data sources, which can be edge devices or servers, store and supply sensing/user data, multimedia files, documents, or AI models. Semantic data are the data that are categorized based on their semantics (i.e., content and utility). Different types of semantic information are embedded in a single data sample, e.g., human faces and behaviors, buildings, weather, context, and locations in the same image. For a specific task, only selected information is useful (e.g., human faces for surveillance). A semantic dataset is identified by its semantics regardless of the physical location, generation mechanism, and communication method. The semantic similarity between two datasets can be measured according to given descriptors such as ``smiley faces", ``German shepherd dogs", and ``hand gestures". Descriptors can be also in the form of multimedia objects such as images, video clips, and speech signals (e.g., the photo of a missing person in Section~\ref{sec: experiments}). From the perspective of task effectiveness, any two semantically similar datasets are identical in their utility. The nodes storing semantically similar data can all supply data to the same requester.
    \item \textbf{Semantic Matching:} SEMDAS can be viewed as a task-oriented semantics-based ``search engine" for edge intelligence. AI algorithms are applied to real-time matching between data and a query in terms of semantic similarity. Specific algorithms are discussed in the next section.
\end{itemize}

\begin{figure*}[t]
    \centering
    \includegraphics[width=1\textwidth]{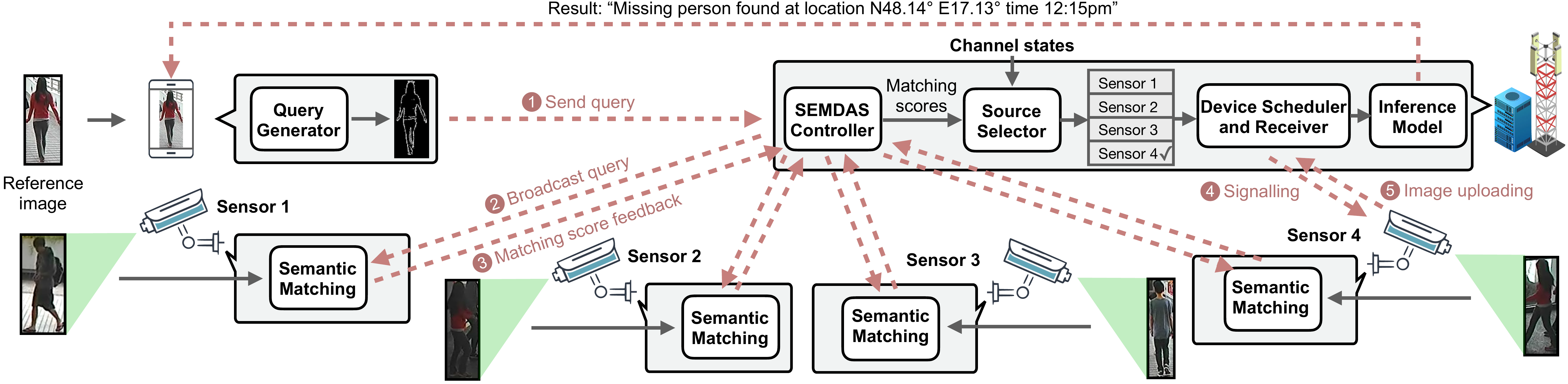}
    \caption{SEMDAS for AI-empowered IoT sensing. }
    \label{fig:iot_sensing}
\end{figure*}
\subsection{SEMDAS Protocol}
A typical SEMDAS protocol comprises the following steps.
\begin{itemize}
    \item \textbf{Service Requesting:} A service requester sends its query (e.g., text or multimedia objects) to the SEMDAS controller. The query is usually a low-dimensional feature vector characterizing the desired data. The requester can be either a user device as in the cases of IoT sensing and edge inference or a server in the case of edge learning. 
    \item \textbf{Semantic Source Selection:} The controller broadcasts the query to all data sources in the network. Each source generates a key of its data and compares it with the query using a suitable semantic matching technique (see Section~\ref{sec: matching}). The operation generates a matching score that is fed back to the controller.
    \item \textbf{Data Uploading:} Using the matching scores, the server selects the best-matched data sources to upload their data. The data is forwarded to the requesting node for processing. Alternatively, the computation can be performed at the controller or servers with the result downloaded to the user. 
\end{itemize}

\subsection{Advantages of SEMDAS}
The proposed SEMDAS framework has the following main advantages. First, SEMDAS represents a scalable solution of data sourcing for edge intelligence. It overcomes the bottleneck of peer-to-peer data transportation by evaluating the semantic similarity between data sources to avoid unnecessary transmission and exploiting the existence of multiple semantically similar sources to cope with communication issues such as long-distance transportation, link disruption, and unreachable sites. Second, data source verification via checking of its semantic content by SEMDAS controller ensures security and data integrity. Third, SEMDAS facilitates mobility management. Unlike 5G connectivity-centric networking, the SEMDAS approach does not require E2E connections. A moving user repeats sending the same query, which can be served by different data sources with semantic similarity. There is no need to maintain a connection to the previous source. Thereby, SEMDAS helps to cope with unfavorable networking conditions, e.g., sparse connectivity, high-speed mobility, and link disruptions.

\section{Semantic Matching Techniques for Edge Intelligence}
\label{sec: matching}
\subsection{Semantic Matching for AI-Empowered IoT Sensing}\label{sec: matching_sensing}
IoT sensing is a new function of 6G that exploits cross-network collaboration between on-device sensors to form a large-scale sensor network for surveillance, localization, tracking, and event detection. The feeding of multi-modal sensing data into edge-AI models endows on the sensors the capabilities of object/event recognition and human behavior detection. Nevertheless, the transportation of high-dimensional sensing data places a heavy burden on the network. The resultant traffic jams can be alleviated by sensor selection based on semantic matching. Relevant applications and techniques are described as follows.

Consider two representative use cases. First, \emph{sensing via crowdsourcing} refers to the involvement of the sensors owned by a group of participating users to collectively perform a sensing task. A specific task of finding a missing person, as illustrated in Fig.~\ref{fig:iot_sensing}, is considered in the experiment in Section~\ref{sec: experiments}. The second use case is \emph{networked perception}. The network function involves multiple devices (e.g., vehicles and robots) cooperating to complete a perception task such as tracking, localization, and object recognition as coordinated by an edge server. The use case of networked perception can be further divided into two sub-cases — \emph{peer-assisted perception} and \emph{multi-view perception}. The former overcomes the issue of a degraded sensor at a device by using matched data collected by a helping device. For instance, an autonomous vehicle with faulty cameras can rely on nearby vehicles to observe the roads and surrounding environment~\cite{Liu2020CVPR}. On the other hand, multi-view perception uses a server to aggregate the 2D observations of multiple camera sensors from different view angles to improve the sensing accuracy (of, for example, object recognition) or reconstruct a 3D object~\cite{Wang2021ICCV}. For instance, the query in the case of peer-assisted perception can comprise a degraded view-image captured by the user's faulty sensor; in the cases of multi-view perception, the query can be generated from an image of a wild hog to locate wild intruders in the city center.  

We propose an efficient semantic matching technique that aims to find an observation close to the query in the semantic space. The semantic space is a feature space where samples (e.g., phrases or images) with similar meanings are clustered and those with different meanings are separated. In practice, the semantic space is created using a projection neural network model that is trained according to the sensing task of interest and deployed to project the query and keys into the space~\cite{Liu2020CVPR}. The matching score of a query-key pair is then obtained with the general attention mechanism, which computes the dot-product or cosine similarity of their projections in the learned semantic space. 
For example, the Google RankBrain system, which powers the Google search engine accessed by billions per day, matches queries and web pages in the semantic space instead of plain keyword-searching. It infers the semantic intent behind the query and identifies web pages that cover the intent, which can be interpreted as projecting both into a semantic space and selecting based on similarity {measurements}. 

The optimal performance of the above semantic matching technique requires the E2E training of two semantic encoders. One is the query generator projecting requests into low-dimensional queries; the other is the key generator computing keys carrying semantic information from sensing observations.

\begin{figure*}[t]
    \centering
    \includegraphics[width=1\textwidth]{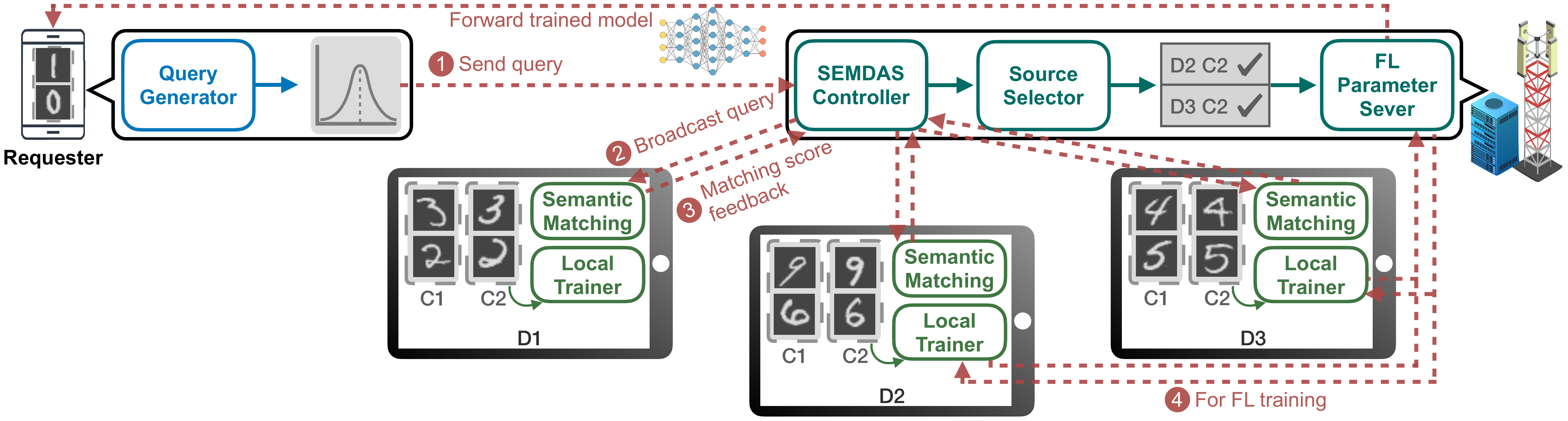}
    \vspace{-2mm}
    \caption{SEMDAS for federated edge learning. }
    \label{fig:edge_learning}
\end{figure*}

\subsection{Semantic Matching for Edge Learning}
\label{sec: matching-learning}
Edge learning refers to the edge-intelligence use case that trains AI models using mobile data distributed in a wireless network. It is envisioned to be a key 6G operation that distills intelligence needed for empowering a wide range of applications. There exist two main approaches. The first is \emph{centralized edge learning} (CEEL) which directly uploads data (or their features) from devices to a server for model training. The other approach, \emph{federated edge learning} (FEEL) is deployed when the data ownership needs to be preserved~\cite{Zhu2020CM}. To this end, FEEL uses a so-called parameter server to download a global model onto devices for updating using local datasets, and then upload and aggregate the local models to update the global model. Implementing the classic iterative algorithm of stochastic gradient descent, the process is repeated until the model converges. Both approaches are confronted with a communication bottleneck as they require uploading of high-dimensional data or model parameters/gradients from potentially many devices. The bottleneck can be alleviated using the SEMDAS approach (see Fig.~\ref{fig:edge_learning}). 

Data at different devices exhibit a high level of heterogeneity. Indiscriminately training a model on all data may be harmful as the inclusion of out-of-domain data can corrupt the model. Then, the purpose of SEMDAS is to find data that match the domain of the learning task. For instance, to learn French-to-English machine translation, the datasets in other languages are irrelevant. As another example, to train a model that recognizes a famous author’s handwriting, not all handwritten texts with the same meaning are useful except for those matching the author’s style. Therefore, the semantic matching in the context of edge learning refers to domain matching between training data and the learning task.

We discuss in the sequel the domain matching techniques for two cases where datasets are characterized by labels or descriptors. The case with data description is relatively simple. The data-task matching can be performed by the SEMDAS controller that receives the data description published by data sources and the task description from the query sent by the data-seeking sender. Then the controller evaluates the score of the matching from a particular dataset to the task by projecting representative data and task descriptors into the semantic space to evaluate their similarity, which is based on the same technique as in Section~\ref{sec: matching_sensing}. The domain matching in the case without dataset description is more challenging. Recently, researchers have observed that different implicit semantic characteristics embedded in a cluster of samples of the same label can cause the cluster to have a nested sub-clustered distribution~\cite{Yuan2022ICLR}. For example, in the MNIST dataset, the handwritten digits of the same label form sub-clusters, each of which corresponds to one writing style~\cite{Yuan2022ICLR}. Therefore, domain matching can be translated to the matching of data distribution and task. To this end, the query is generated to contain statistical information on the data in the desired domain or a representative mini-batch of such data. Then a matching score of a particular dataset is generated by evaluating or approximating its \emph{Kullback-Leibler} distance from the desired distribution specified by the query. The controller coordinates selected devices to upload their data to the requesting server in the case of CEEL or to participate in FEEL.

\begin{figure}[t]
    \centering
    \includegraphics[width=0.48\textwidth]{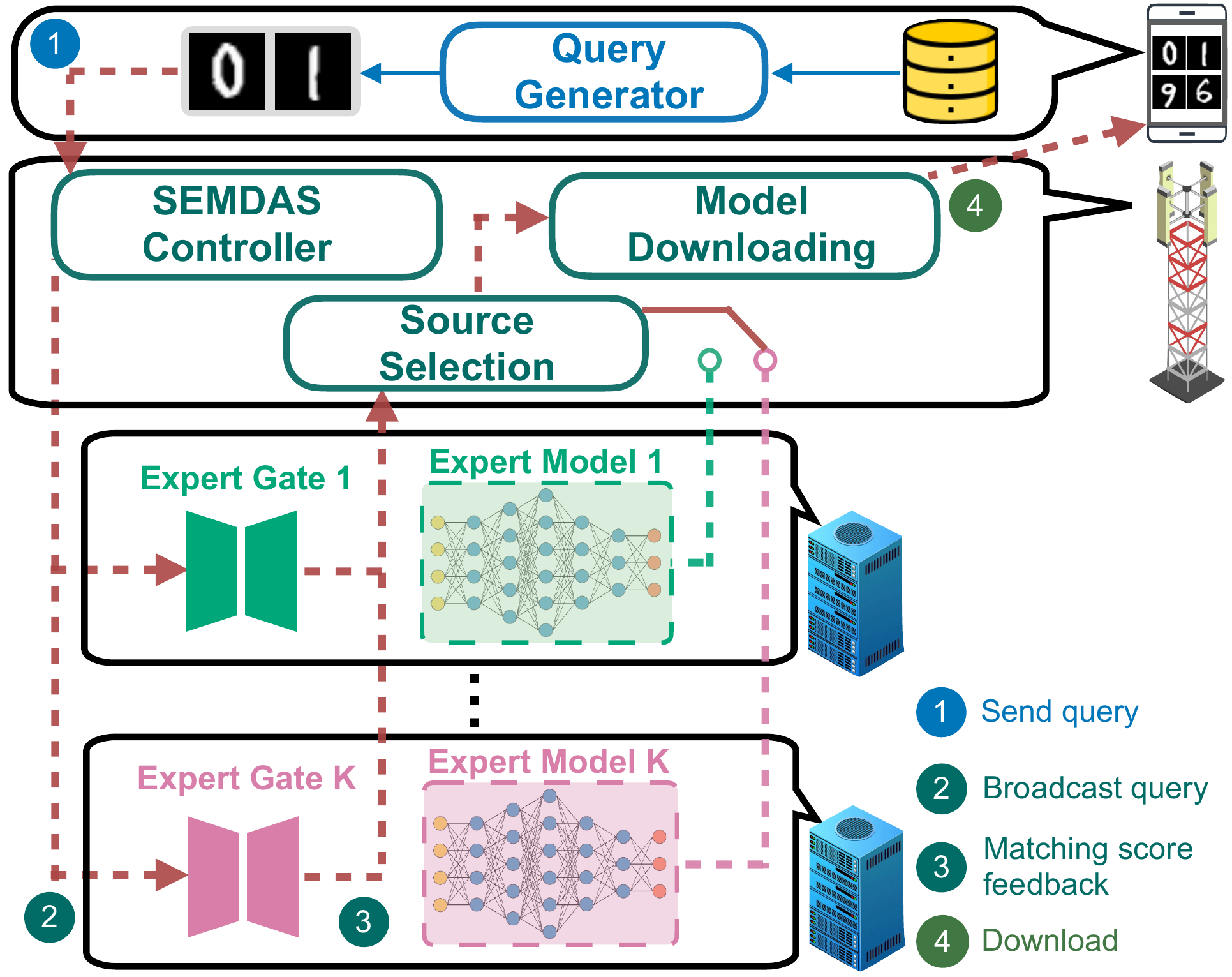}
    \caption{SEMDAS for edge inference with model downloading. }
    \label{fig:model_downloading}
\end{figure}

\subsection{Semantic Matching for Edge Inference}
\label{sec: matching-inference}
Edge inference is a basic operation of 6G edge intelligence that focuses on the efficient provisioning of AI inference capabilities to edge devices. Depending on whether to offload computation to edge servers, edge inference can be implemented in two ways, namely on-device inference and split inference. The former refers to on-demand downloading of AI models to devices for local inference, depicted in Fig.~\ref{fig:model_downloading}. This is suitable for small-to-medium model sizes and provides a faster response speed and better protection of privacy. For the latter, the computation load is split between users and servers. The device sub-model extracts intermediate features from data samples to preserve the data ownership by avoiding sharing raw data. Then the server receives and feeds the uploaded features into the server sub-model to perform prediction. By offloading intensive computation to the server, split inference provides resource-constrained devices access to large-scale deep neural network models, for which on-device deployment is infeasible. There exist a practically infinite number of models stored in the network. They can be differentiated in many dimensions, e.g., size, context, task, performance, and topology~\cite{Huang2022arxiv}. For instance, an object recognition model can either have a convolutional or recursive topology and can be trained for an urban, rural, or indoor environment. In the context of edge inference, a data source refers to a node (device or server) that stores a particular sharable model (i.e., its parameters and topology information). Then the SEMDAS problem is to find data sources whose models meet the requester’s requirements in terms of inference accuracy, complexity, and storage. 

We discuss the semantic matching techniques for edge inference as follows. First, the technique for the case with available model description is similar to those for sensing and edge learning, for which matching leverages semantic space. Next, we focus on the case where the model description is either unavailable or insufficient for semantic matching. The proposed techniques for this purpose require a requesting device to generate a query containing test samples of local data (e.g., sensor observations). Then using the query, the SEMDAS controller performs semantic matching via either of the following two techniques. The first is \emph{expert gateway}. Assume that the controller has access to a library of popular AI models called expert models. Each expert has a lightweight autoencoder pair of the encoder and the decoder apart from the inference model itself. The controller consults each expert by providing it with the query. In response, the expert uses its autoencoder to extract features from the test samples and generate reconstruction errors~\cite{Aljundi2017CVPR}. The controller selects the expert with the minimum errors (i.e., the one best matching the user's task) to download its model. As the second technique called \emph{server polling}, if no matched model can be found among expert models, the SEMDAS controller broadcasts the user’s query to all available servers in the network. Each server tests the matching level of its model by using it to perform inference on the test samples in the query and evaluate the generated confidence scores. The model with the highest overall score is deemed semantically matching the query. Then the hosting server is associated with the requesting user to provide the inference service in the case of split inference. In the case of model downloading, the model is retrieved and forwarded to the user.

\section{Task-Oriented Communications for SEMDAS}
\label{sec: communications}
Task-oriented communication techniques for SEMDAS are designed using an E2E performance metric such as IoT sensing accuracy, convergence speed for FEEL, and prediction error for edge inference. In this section, we introduce this new area by proposing new design principles and discussing several research opportunities. 

\subsection{Tradeoffs and Optimal Designs}
There exist two unique tradeoffs in wireless communications for SEMDAS. They allow optimization of the edge-intelligence system and techniques. 

The first tradeoff, called the \emph{query-data tradeoff}, is due to two opposite trends.  On one hand, shortening the query vector reduces the overhead in broadcasting to all data sources.  On the other hand, less information contained in queries can lead to less accurate semantic matching and hence more uploaded data that increase the uplink overhead.  The opposite trend also holds, inducing the said tradeoff between the query and data overhead. Wireless systems for SEMDAS can be optimized based on the tradeoff. For instance, the uplink-downlink division of bandwidth or time can be optimized for an E2E performance metric. Furthermore, under a query-rate constraint, the query can be designed to maximize the overall semantic matching level at data sources or to minimize the communication overhead for data uploading for a given matching level.

The second tradeoff, called the \emph{privacy-accuracy tradeoff}, results from the fact that placing more details in the query (e.g., the number of features of a missing person) reveals more private information about the requestor (e.g., the person's identity) or his/her task but improving the semantic-matching accuracy and hence the communication efficiency, and vice versa. The tradeoff is of practical interest as in many practical scenarios, the query contains private information and its broadcasting over the network can compromise the requester's privacy. The consideration of the tradeoff leads to the need for privacy-preserving communication designs. In particular, under a privacy constraint on the requester, the query generation can be integrated with adaptive transmission (e.g., coding, power control, and beamforming) to maximize the system efficiency.

\subsection{Joint Semantic-and-Channel Matching (JSCM)}
Following a rate-centric approach, selecting data sources merely based on their channel states can result in a mismatch between the sourced data and the computing task. Nevertheless, the consideration of only semantics-matching in the selection may lead to unreliable communication links. Therefore, for supporting SEMDAS applications in wireless systems, it is important to balance both aspects in choosing data sources, which is called JSCM. Redesigning traditional rate-centric wireless techniques to feature JSCM creates many research opportunities. Several selected ones are described as follows. 
\begin{itemize}
    \item \textbf{Multi-Access:} JSCM-based distributed selection of data sources for uploading over a multi-access channel can be implemented using a threshold on their semantic matching scores and another on their channel gains. The thresholds can be jointly designed for the dual objectives of sourcing sufficient relevant data and at the same time regulating the number of accessing devices to avoid frequent packet collisions or insufficiency of radio resources. 
    \item \textbf{\emph{Over-the-Air Computing} (AirComp):} AirComp exploits the waveform superposition property of a multi-access channel to realize over-the-air aggregation of views and local models/stochastic gradients in the use cases of sensing and FEEL, respectively~\cite{Amiri2020TSP}. The bottleneck of suppressing AirComp error lies in those channels of transmitters with unfavorable conditions. JSCM-based device selection can be designed to balance the suppression of AirComp error and sourcing sufficient data to optimize the E2E system performance. 
    \item \textbf{\emph{Radio Resource Management} (RRM):} Traditional RRM techniques such as sub-channel allocation, power control, and scheduling have been designed using a rate-centric metric. To be task-oriented for SEMDAS, these techniques should be redesigned to prioritize devices not only based on their channel states but also their semantic matching scores. 
    \item \textbf{Random Beamforming:} The existing random beamforming schemes exploit multiuser channel diversity to maximize the data rate by selecting among many users those with beam-aligned channel vectors and large channel gains. Such schemes can be modified to simultaneously exploit the diversity in both channel and semantics over multiple data sources. 
\end{itemize}

\section{Experiments}
\label{sec: experiments}
\subsection{Experiment Design}
We demonstrate SEMDAS using the sensing application of finding missing people (e.g., a child or an elderly) via a network of wirelessly connected surveillance cameras. The underpinning operation, technically known as \emph{re-identification} (ReID), attempts to associate camera views, which are captured in different occasions, with the same person that is specified by the reference photo in the query sent by the person’s families. The real-time location information shared by the matched cameras would help locate the person. To implement semantic matching, we adopt the LightMBN model trained on the well-known CUHK-03 training set~\cite{Herzog2021ICIP} for semantic feature extraction and use the matching function discussed in Section~\ref{sec: matching_sensing}. Thereby, the query photo and each camera view are compressed into translated into  3584-by-1 feature vectors for the purpose of semantic matching. For communication, the query broadcast to sensors is at 32-bit resolution per dimension; the matched sensors upload their raw data (i.e., photos) for their views with an Acknowledge of “person found”. We simulate a network of 20 cameras and generate the query and camera views by drawing samples from the CUHK-03 dataset. In each trial, among 20 camera views, 4 contain the target identity while the others are associated with different persons. In total, the experiment contains 1229 random trials. The channels from sensors to the SEMDAS controller are modeled as i.i.d. Rayleigh fading and being orthogonal with a total uplink bandwidth of 5 MHz. The performance metric, called missing rate, is defined as the probability of ReID failure, namely that none of the matched cameras views actually contains the target person. The missing rate is evaluated as a function of average uplink communication load (in Mbits) and average uplink communication latency (in milliseconds) per request. We consider the following JSCM and benchmarking schemes in experiments. 
\begin{itemize}
    \item (Proposed) JSCM: A given number of matched sensors are selected using the criterion of maximizing the weighted sum of semantic matching score and communication rate, where their weights are optimized numerically as 1 and 0.09, respectively.
    \item \emph{Best-semantics selection} (BSS) based on the criterion of the maximum semantic matching score.
    \item \emph{Best-channel selection} (BCS) based on the criterion of the maximum communication rate.
    \item \emph{Random selection} (RS) of uploading sensors.
\end{itemize}
\begin{figure}[t]
    \centering
    \includegraphics[scale=0.45]{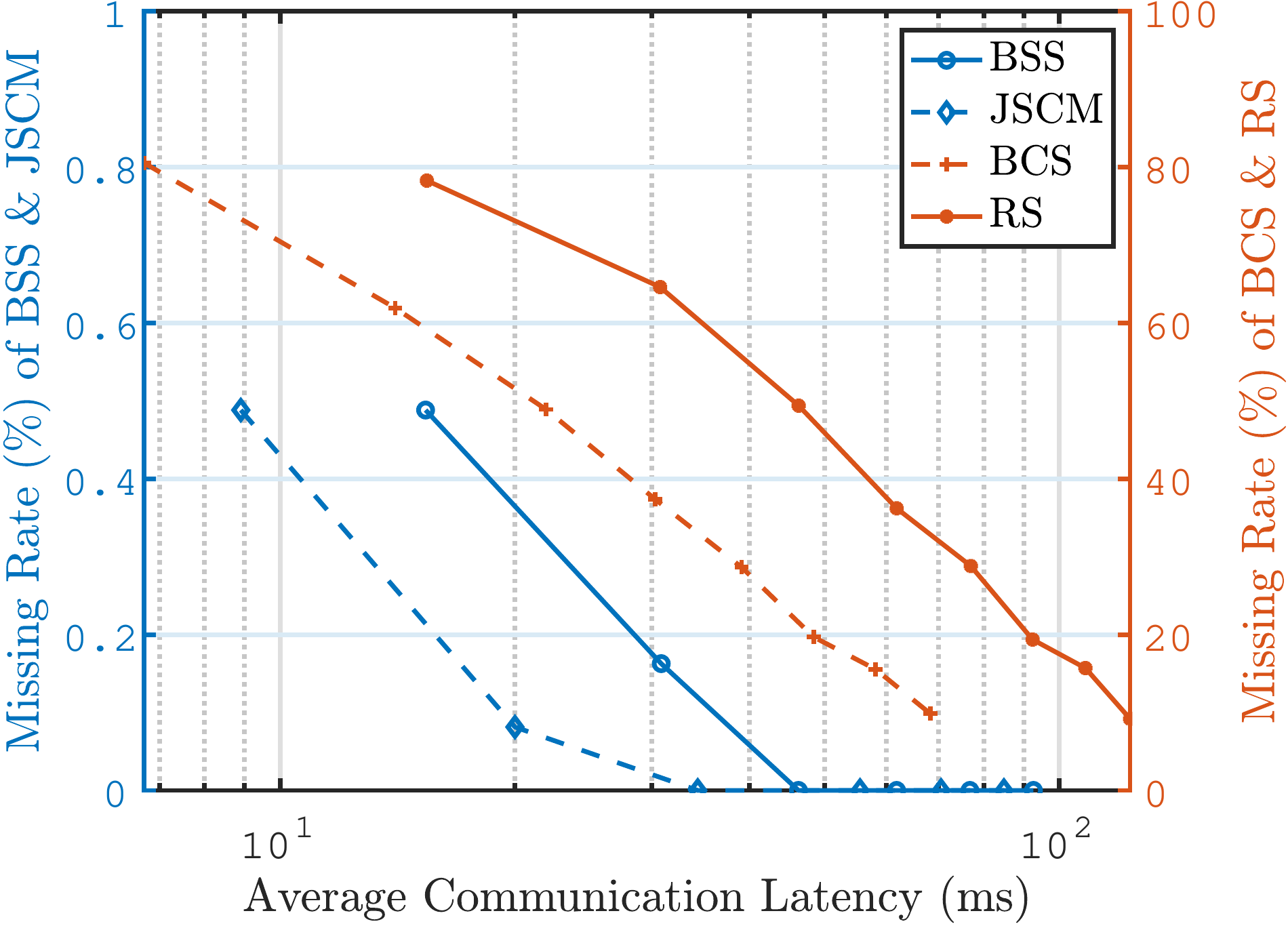}
    \vspace{-1mm}
    \caption{The missing rate versus average communication latency for the proposed JSCM and three benchmarking schemes — BSS, BCS and RS. }
    \label{fig:5}
\end{figure}
\subsection{Performance Evaluation}
The performance of JSCM is compared with that of BSS, BCS and RS. Fig.~\ref{fig:5} depicts the curves of missing rate versus communication latency that grows as the number of selected sensors increases. We can observe that BCS, a rate-centric scheme, and RS both have unacceptable performance due to their lack of semantic awareness. On the other hand, with such awareness, the missing rates of the JSCM and BSS schemes achieve much lower missing rates than the preceding schemes, for example, more than 2-order magnitude lower at 30-ms latency. Between JSCM and BSS, the proposed design significantly outperforms the latter. This demonstrates JSCM being a promising solution and the need for task-oriented wireless design.

\section{Concluding Remarks}
\label{sec: conclusion}
We have proposed the SEMDAS framework to solve the problem of communication bottleneck of 6G systems caused by data sourcing in edge-intelligence use cases. Its basic principle is to transport only data whose semantics match the computing tasks so as to avoid redundant network traffic due to transmission of irrelevant data. Based on the principle, a comprehensive framework has been presented that comprises the architecture, protocol, learning-based semantic matching techniques, as well as new design principles for task-oriented wireless techniques. At the high level, SEMDAS represents a contribution to the ongoing development of 6G semantic communication systems. In particular, the new framework points to the new direction of revolutionizing the wireless network architecture to enable highly efficient edge-intelligence operations and applications.

\bibliographystyle{IEEEtran}

\begin{IEEEbiographynophoto}{Kaibin Huang} [Fellow, IEEE] is a Professor at the Department of Electrical and Electronic Engineering, The University of Hong Kong, Hong Kong. His research interests include mobile edge computing, edge AI, and 6G systems. 
\end{IEEEbiographynophoto}

\begin{IEEEbiographynophoto}{Qiao Lan} received his B.Eng. degree from Southern University of Science and Technology (SUSTech), Shenzhen, in 2019. He is now pursuing a Ph.D. degree with The University of Hong Kong.
\end{IEEEbiographynophoto}

\begin{IEEEbiographynophoto}{Zhiyan Liu} received his B.Eng. degree from Tsinghua University, Beijing, in 2021. He is now pursuing a Ph.D. degree with The University of Hong Kong.
\end{IEEEbiographynophoto}

\begin{IEEEbiographynophoto}{Lin Yang} received the Ph.D. degree from Hong Kong University of Science and Technology and is now a senior researcher in Huawei Noah's Ark Lab.
\end{IEEEbiographynophoto}

\end{document}